\documentclass{epl}

\title{Electrical conductance of a 2D packing of metallic beads 
under thermal perturbation}
\author{D. Bonamy\inst{1} \and L. Laurent\inst{1} \and Ph. Claudin\inst{1,2} 
\and 
J.-Ph. Bouchaud\inst{1} \and F. Daviaud\inst{1}}
\institute{
  \inst{1} Service de Physique de l'Etat Condens\'e - CEA Saclay, 91191 Gif-sur-Yvette 
Cedex, France\\
  \inst{2} Technion - Israel Institute of Technology - Physics department, Haifa 32000, 
Israel
}

\begin{document}

\maketitle

\begin{abstract}
Electrical conductivity measurements on a 2D packing of metallic beads have 
been performed to study internal rearrangements in weakly perturbed 
granular materials. Small thermal perturbations lead to large non gaussian 
conductance fluctuations. These fluctuations are found to be intermittent in 
time, with peaks gathered in bursts. The distributions of the waiting time 
$\Delta t$ between two peaks is found to be a power-law $\Delta 
t^{-(1+\alpha_t)}$ inside bursts. The exponent $\alpha_t$ 
is independent of the bead network, the intensity of perturbation and 
external stress. These bursts are interpreted as the signature of individual 
bead creep rather than collective vault reorganisations. We propose a simple
model giving $\alpha_t=1-\zeta$, where $\zeta$ is
the roughness exponent of the surface of the beads.
\end{abstract}

\section{Introduction}
Granular materials present interesting and unusual properties~\cite{b.1}. For 
example, photoelastic visualizations of static confined granular packing 
under 
external stress have provided evidence of large inhomogeneities in the 
distribution of contact forces between grains~\cite{b.2,b.3}, giving rise to 
strong force chains (vaults or arches) extending on a scale much larger than 
the size of 
an individual grain. Recent experiments have revealed the sensitivity of 
sound 
transmission in granular media to small perturbations (such as thermal 
expansion 
of the grains)~\cite{b.4}. This sensitivity  has been interpreted 
using a simple theoretical model of vault formation, the `Scalar Arching 
Model'~\cite{b.5}. This model predicts that large rearrangements of the force 
chains 
network can indeed occur, even for small external perturbations. This 
sensitivity tends to disappear when the 
media is subjected to a strong external stress~\cite{b.6}.

However, both photoelastic visualization and sound transmission techniques 
require large external stresses and can not give information on a weakly 
confined granular media where interesting signatures of {\it 
fragility}~\cite{b.7} can be expected. We propose here to study this 
fragility through conductivity measurements. Experiments on the electrical 
properties of granular media have been performed in the past : Branly 
reported ~\cite{b.8} the influence of an electromagnetic wave on the 
electrical resistance of a granular packing. Giraud et al. ~\cite{b.9}, 
Marion et al.~\cite{b.10} and Gervois et al.~\cite{b.11} have performed 
measurements of the electrical conductivity in order to reveal the influence 
of the contact network on the mechanical properties of the media. More 
recently, Vandewalle et al. ~\cite{b.12} reported power-law distributions of 
conductivity changes, and argued that this was a signature of large 
rearrangments of stress paths.

The aim of this work is to investigate whether force chain rearrangements can be observed 
through the variation of the electrical conductance of a 2D bead packings 
under small thermal perturbations. Large intermittent fluctuations are indeed observed. They appear in bursts 
that can be characterized via a statistical analysis of the waiting time separating two fluctuation peaks. The 
microscopic origin of these electrical fluctuations is then investigated. We argue that most of the interesting 
statistical features are not related to large force chains 
rearrangements but to individual microcontacts between two beads that 
rearrange.
The observed non trivial statistics might be related to the self-affine
roughness of the surface of the beads. 

\section{Experiment}
\begin{figure}
\onefigure[width=14cm]{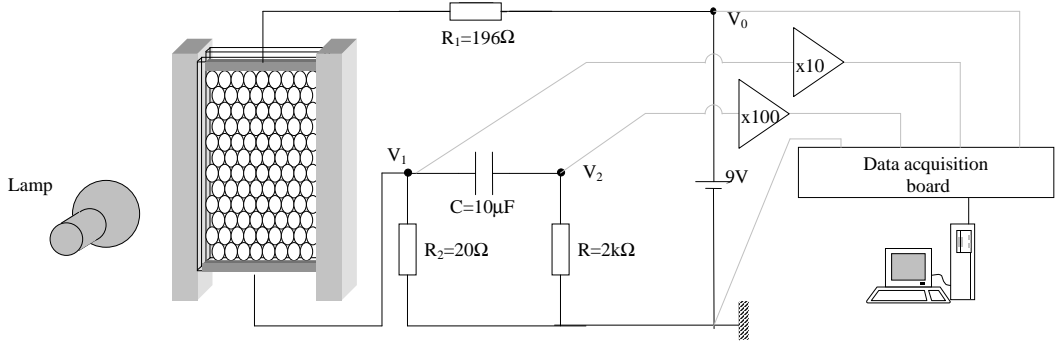}
\caption{experimental set-up}
\label{f.1}
\end{figure}

The experimental set-up is illustrated on fig.~\ref{f.1}. It consists mainly 
of 
a 2D packing of stainless steel beads of diameter $a=6\un{mm} \pm 10\un{\mu 
m}$ confined between two Plexiglas plates (thickness $2\un{cm}$). The use of metallic beads of 
millimetric size allows one to have a good control of geometrical and 
mechanical properties. We can also change the number of grains in the sample 
to separate local behaviour from collective behaviour.
Beads are carefully arranged to form triangular compact packing. This 
situation 
of maximal compacity can easily be reproduced before each experiment. Even if 
the piling can be viewed as regular, the weak polydispersity and the solid 
friction between grains is sufficient to entail disorder in the network of 
contact forces~\cite{b.13}. The size of the packing is $W \times H$ where $W$ 
is its width and $H$ its height in bead size units. Thermal perturbations are 
induced by a $75\un{W}$ lamp standing at a distance $d=10\un{cm}$. The lamp 
increases the temperature of the beads and of the Plexiglas plates by 
typically $3\un{^\circ C}$. This leads to an expansion of both the beads and 
the
plates: the thermal expansivity for beads is $2 \ 10^{-5}\un{K^{-1}}$ and 
$7\ 10^{-5}\un{K^{-1}}$ for Plexiglas. The bead network can also be vibrated via a buzzer placed against the 
Plexiglas plate. The packing is connected to 
a $9\un{V}$ battery. A resistor $R_{1}$ in series insures that the current crossing 
a contact between any couple of beads is much smaller than $40\un{mA}$. 
Preliminary studies performed on 2 beads indeed show  that, as long as this 
current is below $40\un{mA}$, micro-welding is prevented and the contacts 
behave reversibly like an ohmic resistance. Voltages $V_{0}$, 
$V_{1}$ and $V_{2}$ (see fig.~\ref{f.1}) are recorded via a 12 bytes AD 
converter: $V_{0}$ is the emf of the battery, $V_{1}$ is proportional to the 
current crossing the packing and $V_{2}$ represents the fast fluctuating part 
(faster than $\tau = RC < 0.02\un{s}$) of $V_{1}$. The following quantities 
can then be deduced:
\begin{itemize}
\item the conductance $g_p$ of the packing. For a $33 \times 52$ vibrated 
packing, $g_p \simeq 0.02\un{\Omega^{-1}}$.
\item the fluctuation of conductance $\Delta g_p$ defined as the variation of 
$g_p$ between 2 acquisition points. Typically, $\Delta g_p$ reaches values up 
to $5.10^{-3}\un{\Omega^{-1}}$ to be compared to the detection threshold 
discussed below, which is typically around $5.10^{-6}\un{\Omega^{-1}}$.
\end{itemize}

A series of experiments was performed with the bead packing replaced by a 
$50\un{\Omega}$ resistor in order to determine the level of noise of the 
acquisition system. This is mainly dominated by digital errors on $V_{2}$ 
equal to $5\un{\mu V}$. A threshold is then defined: $\Delta g_p$ is 
considered as relevant when the corresponding $\Delta V_2$ is superior to 
$50\un{\mu V}$. Sampling rates $f$ from $0.05\un{Hz}$ up to $10\un{kHz}$ have 
been tried: $f=500\un{Hz}$ is sufficient to separate most successive events 
(cf fig.~\ref{f.2}). For each experiment, data is recorded during a period 
of 20 minutes. The procedure is the following : (i) the packing is vibrated, 
(ii) the acquisition is started, (iii) the light is turned on 1 minute after (turned off in experiments with packing 
prepared with light switched on).

\section{Results}
\begin{figure}
\onefigure[width=14cm, height=5.0cm]{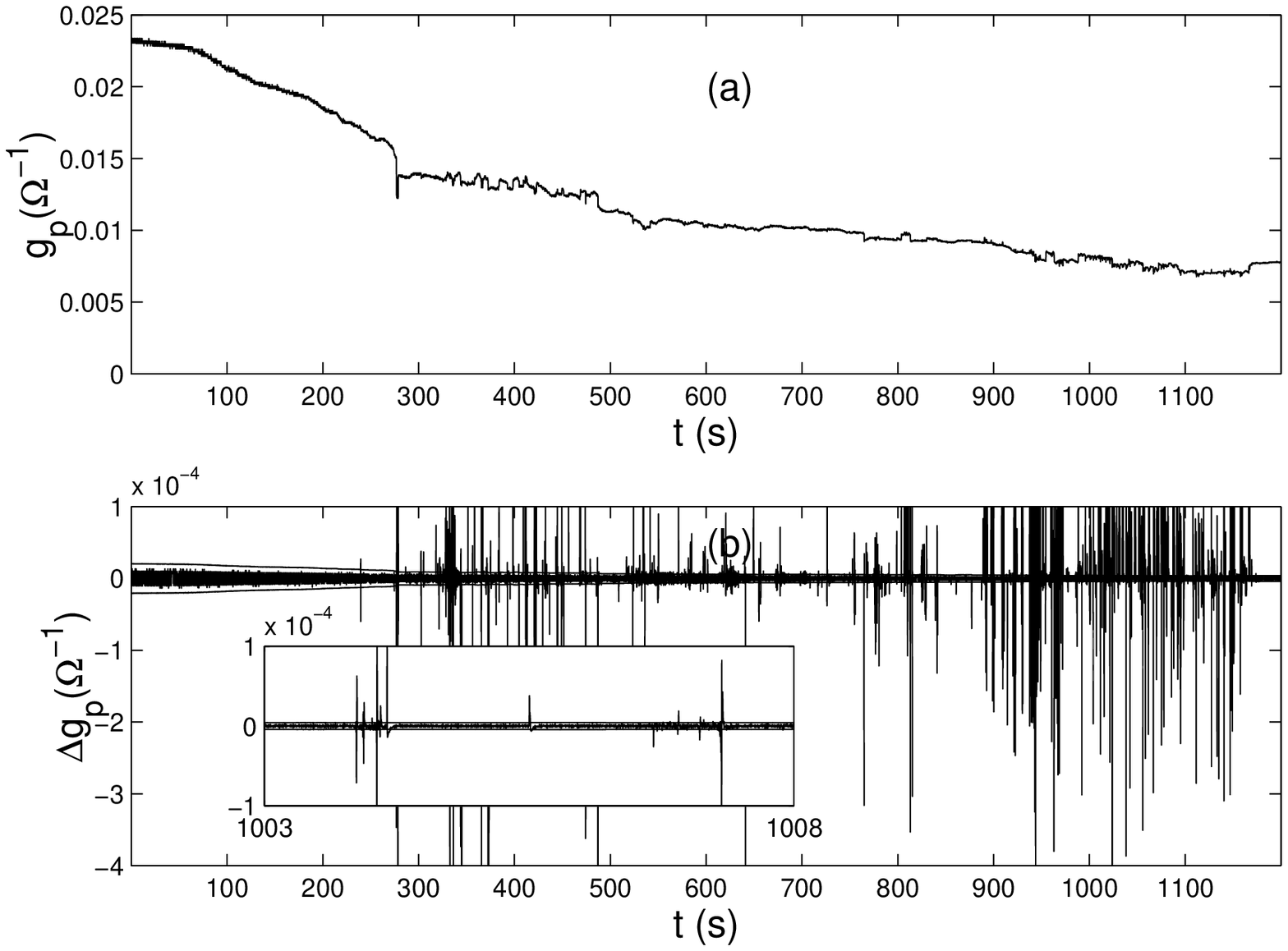}
\onefigure[width=14cm, height=5.0cm]{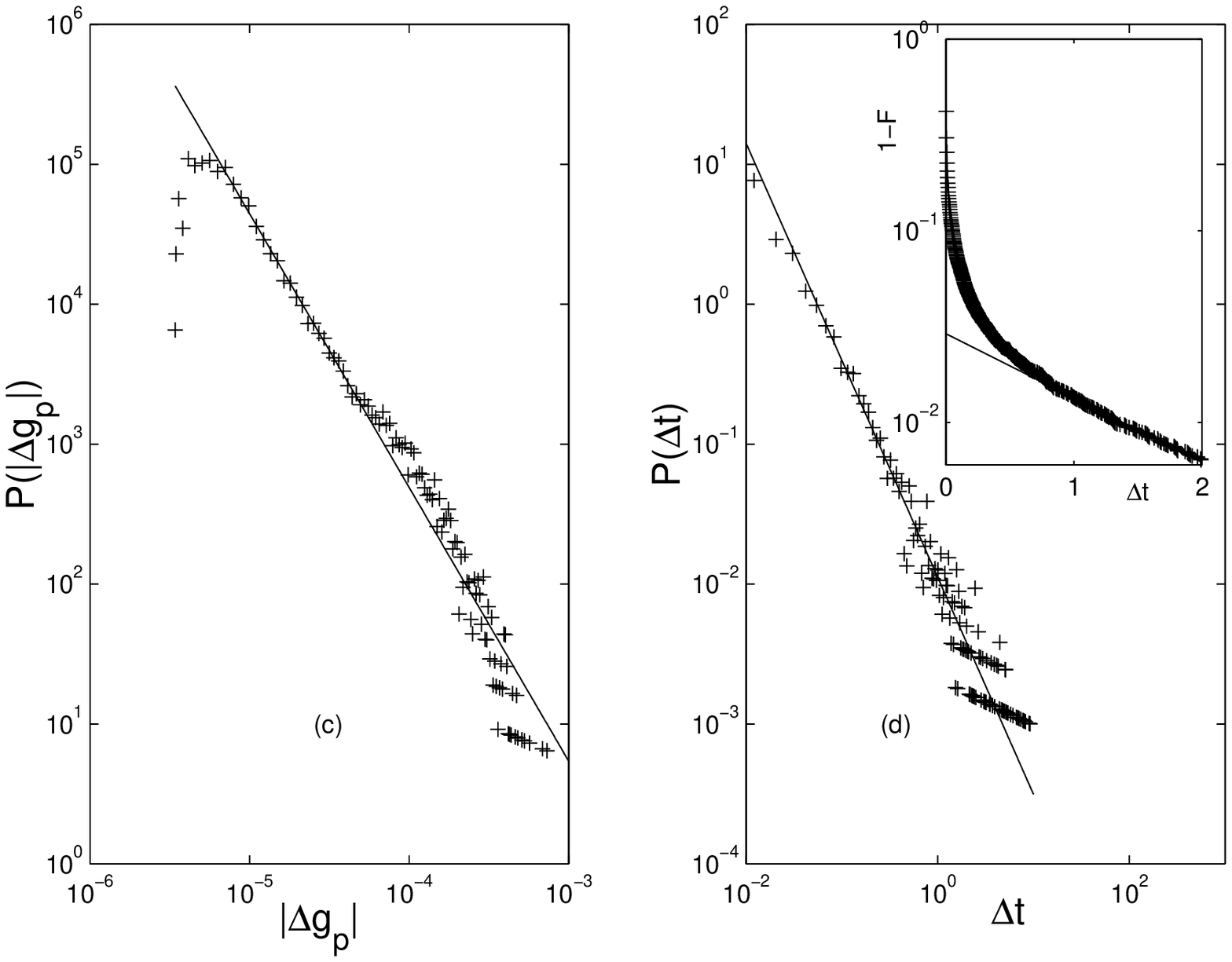}  
\caption{Analysis of $\Delta g_p$ for a $33\times 52$ regular packing heated 
with the lamp. (a) Temporal evolution of the $g_p$. (b) Temporal evolution of $\Delta g_p$. Inset, 
expanded view on a few seconds scale (c) Log-log plot of the distribution $P(|\Delta g_p|)$. The straight line is a 
power law fit using Eq.~\ref{e.1} with $\alpha_g=1$.(d) Log-log plot of the probability density function 
$P(\Delta t)$. The straight line is a power law fit using Eq.~\ref{e.2}, with $\alpha_t=0.61$. The inset shows the 
cumulative distribution which decays asymptotically as an exponential.} 
\label{f.2}
\end{figure}

A packing, initially vibrated, keeps a constant conductance as long as its temperature does not change. As soon 
as it is thermally perturbed (light turned on or off), $g_p$ starts to decrease and to fluctuate. Let us focus firstly 
on the fluctuations $\Delta g_p$ for a typical experiment (see fig.~\ref{f.2}) on a $33 \times 52$ packing. 
Fluctuation peaks appear in an intermittent way. The distribution of $\Delta g_p$ is to a good approximation 
symmetrical and can be described (at least for small enough $|\Delta g_p|$) by the following power-law:
\begin{equation}
P(|\Delta g_p|) \sim |\Delta g_p|^{-(1+\alpha_{g})}
\label{e.1}
\end{equation}
For this experiment, $\alpha_{g} \simeq 1$, in agreement with the results 
reported in ~\cite{b.12}. For large $|\Delta g_p|$, $P(|\Delta g_p|)$ departs 
from this power-law behaviour. One can also study, for the same experiment, 
the temporal structure of the apparition of these peaks. The expanded view 
fig.~\ref{f.2}b of $\Delta g_{p}(t)$ presents a burst structure, which can be 
characterised by the distribution of the waiting time $\Delta t$ between two 
successive events (independently of their sign). Two distinct regimes can be observed in fig.~\ref{f.2}d: 
\begin{itemize}
\item The short $\Delta t$ part of the probability density $P(\Delta t)$ 
(corresponding to events within the same burst) decays as a power law with 
$\Delta t$:
\begin{equation}
P(\Delta t) \sim \Delta t^{-(1+\alpha_{t})}
\label{e.2}
\end{equation}
The exponent is found to be $\alpha_{t} \simeq 0.6$.
\item for larger $\Delta t$, $P(\Delta t)$ again departs from this power-law
behaviour. The cumulative distribution function $F(\Delta t)$ is less noisy 
for large $\Delta t$ and shown in the inset of fig.~\ref{f.2}d. One clearly 
sees that $F$ decays exponentially for large $\Delta t$, which is the 
signature 
of a Poisson flux. The bursts themselves therefore appear as completely 
independent events, whereas the events inside bursts are strongly 
non-Poissonian.
\end{itemize}

\begin{figure}
\onefigure[width=14cm]{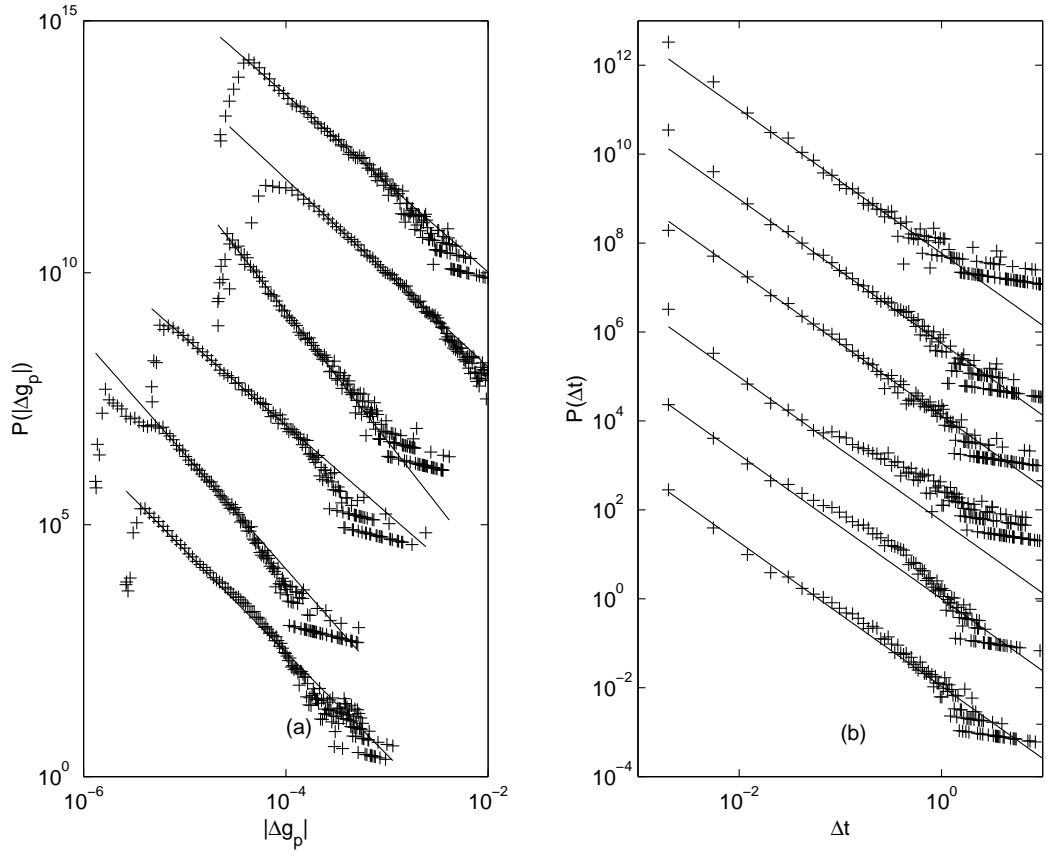} 
\caption{Log-log plot of (a) $P(|\Delta g_p|$ and (b) $P(\Delta t)$ for 
packing 
(1), (2), (3), (4), (5), and (6) from bottom to top. Data have been shifted 
for clarity.
The straight lines are power-law fits. The exponent $\alpha_t$ has been fixed 
for all $P(\Delta t)$ to $\alpha_t=0.6$ minimising errors on the fit for the 
three bead packing (6). $\alpha_g$ is equal, from bottom to top, to 1.0, 1.1, 
0.8, 1.6, 0.9, and 0.8 respectively.}
\label{f.3}
\end{figure}

In order to determine the origin of these electrical fluctuations, several 
experiments have been performed with different packing: (1) regular $33 
\times 
52$ packing, (2) disordered packing of 1690 beads, (3) regular $33 \times 32$ 
packing, (4) regular $33 \times 16$ packing, (5) regular $33 \times 2$ 
packing. 
Finally, (6) is a regular $33 \times 2$ packing where all metallic beads have 
been replaced by insulating glass beads except 3 beads forming a triangle in 
the middle. For all these experiments, $P(|\Delta g_p|)$ decays as a 
power-law but the exponent $\alpha_g$ depends on the bead network geometry, 
and lies between
$0.8$ and $1.6$ (see fig.~\ref{f.3}a). On the contrary, the power-law decay 
of $P(\Delta t)$ does not depend on the packing, and leads to $\alpha_t \sim 
0.6$ (see fig.~\ref{f.3}b). For the packing of 3 beads, the power-law 
behaviour actually extends over the three decades of the time distribution: 
only few bursts are present. The exponent is found to be $\alpha_t=0.6 \pm 
0.05$. When the number of beads increases, the value of $\Delta t$ above which 
$P(\Delta t)$ starts to deviate from a power-law decreases. It can be interpreted by the increase of independent 
burst number making 
the inter-burst waiting time mix more and more with the intra-burst 
statistics.  The intensity of the thermal perturbation has been changed by 
modifying the distance of the lamp ($d=5\un{cm}$, $10\un{cm}$ and 
$20\un{cm}$) without affecting the value of $\alpha_t$. Moreover, for 
packing (5), the weight of the upper electrode has been varied 
($M=108\un{g}$, $M=206\un{g}$, $M=304\un{g}$ and $M=404\un{g}$). The number 
of events decreases, but $P(\Delta t)$ remains a power-law with the same 
exponent $\alpha_t=0.6$.

\section{Discussion}

The origin of these non trivial electric fluctuations should be found in the 
geometry of electrical paths. A natural idea would be to relate these to 
force chains rearrangements. Indeed the Scalar Arching Model (SAM) 
introduced by two of us~\cite{b.5} suggests the existence of large force 
chains rearrangements in a packing subjected to small thermal perturbations. 
The SAM predicts a broad distribution of the apparent weight fluctuation 
$\Delta W_{a}$ measured at the bottom of the packing. This broad distribution 
of $\Delta W_{a}$ could in principle be related to the observed broad distribution 
of $\Delta g_p$. Moreover, the 
model also predicts avalanching contact reorganisations whose electrical 
signature could indeed be the bursts of fluctuations peaks seen in the thermally
perturbed packing, each $\Delta g_p$ peak corresponding to the creation or 
the breakdown of a contact somewhere in the packing. However, in this 
scenario, the intra-burst structure should become less and less pronounced 
when the number of beads decreases, and should disappear completely for a 
packing of 3 beads. This is clearly not the case (see fig.~\ref{f.3}b).

The fact that the power-law nature of the distributions is independent of the 
geometry of the packing, and still holds for three beads,
suggests that the origin of these electrical fluctuations is local. The 
direct imaging of the surface of the beads with an Atomic Force Microscope 
allowed us to measure the average 
roughness $R$ of the beads. We have found $R=90\un{nm}$, to be compared to 
Hertzian deformation $\delta$ in the packing. For contact forces on the order 
of $10^{-2}$ Newton, $\delta \simeq 10\un{nm}$ which is smaller than $R$. The 
electrical contact between 2 beads is thus completely dominated by the 
surface roughness. We propose then the following scenario to explain the 
observed behaviour of $g_p$ and $\Delta g_p$ in our experiment: when 
the packing is vibrated, the surface roughness of the two surfaces in contact 
`adapt' to one another and the effective contact surface is high. Therefore, 
$g_p$ is maximal. As soon as the packing 
is heated, beads expand on the order of $0.5\un{\mu m}$. This thermal 
expansion 
is hindered by the roughness induced friction between beads. Therefore, most 
of
the time the contact between two beads does not change, until the accumulated
stress due to the thermal expansion is sufficient to `unpin' the rough
surfaces and make them slip over each other. This leads to the burst structure
of the signal. Inside each burst, individual peaks of $\Delta g_p$ 
correspond to the creation or the loss of a micro-contact between the two
slipping beads. These micro-contact avalanches globally lead to a smaller 
effective contact surface, thereby making $g_p$ smaller. 

Simple models of pinning \cite{b.14,b.15} can lead to a power-law distribution of the `trapping' time between 
successive events; however, in these models the exponent $\alpha_t$ is found to depend on the external driving 
force. The fact that $\alpha_t$ is independent of both the strength of the perturbation and of the external stress 
suggests a purely geometrical interpretation: if the bead surface can be considered to be self-affine with 
a Hurst exponent $\zeta$, the distance $\Delta \ell$ separating two successive 
`spikes' in a given direction is power-law distributed with an exponent 
$\alpha_\ell=1-\zeta$ on the interval $[a_1,a_2]$ where $a_1$ is a microscopic 
length below which the surface is flat, and $a_2$ is at most the diameter of 
an Hertzian contact: $a_2 \simeq 1\un{\mu m}$. The typical speed $V$ with 
which a bead thermally expands is $1\un{\mu m}$ per second. If this speed is 
constant in time, the time between 
successive micro-contact closing/opening is therefore a power law 
distribution with an exponent $\alpha_t=1-\zeta$ in the interval 
$[a_1/V,a_2/V=1\un{s}]$, which is precisely the experimental interval of time 
scales. We have determined $\zeta$ from AFM frames via the moving average 
method~\cite{b.16}: the moving average of a series $z(x)$ on a length 
interval $L$ is defined as :
\begin{equation}
\overline z(x)=\frac{1}{L}\sum_{i=0}^{L-1}z(x+i)
\label{e.3}
\end{equation}
For a self affine signal with an Hurst exponent of $\zeta$, the density 
$\rho$ of crossing point between two moving averages on length scales $L_1$ 
and $L_2>L_1$ is found to scale as:
\begin{equation}
\rho \sim \frac{1}{L_2}[(\Delta L)(1-\Delta L)]^{\zeta-1} 
\label{e.4}
\end{equation}
where $\Delta L=(L_2-L_1)/L_2$.
 
\begin{figure}
\twoimages[width=7cm]{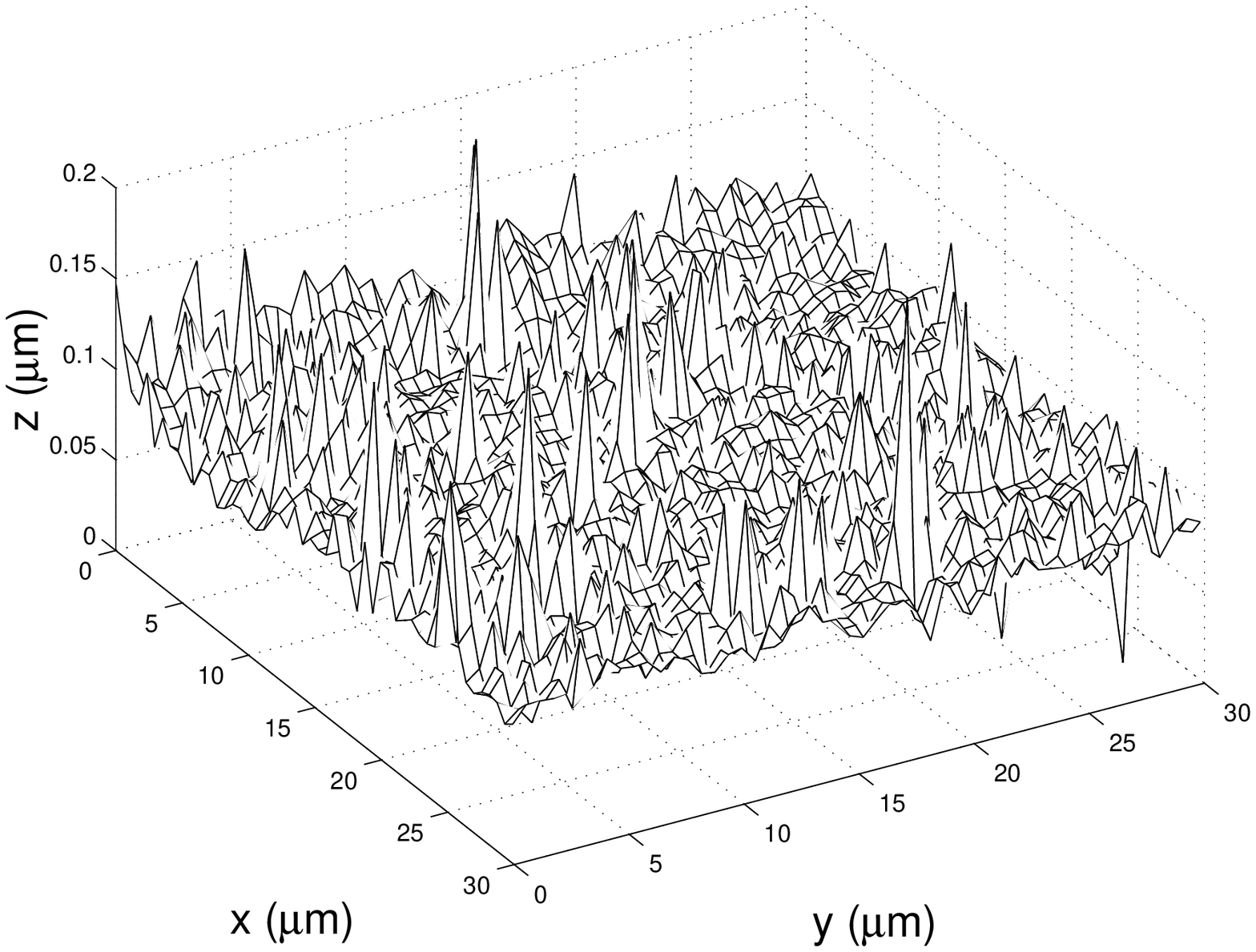}{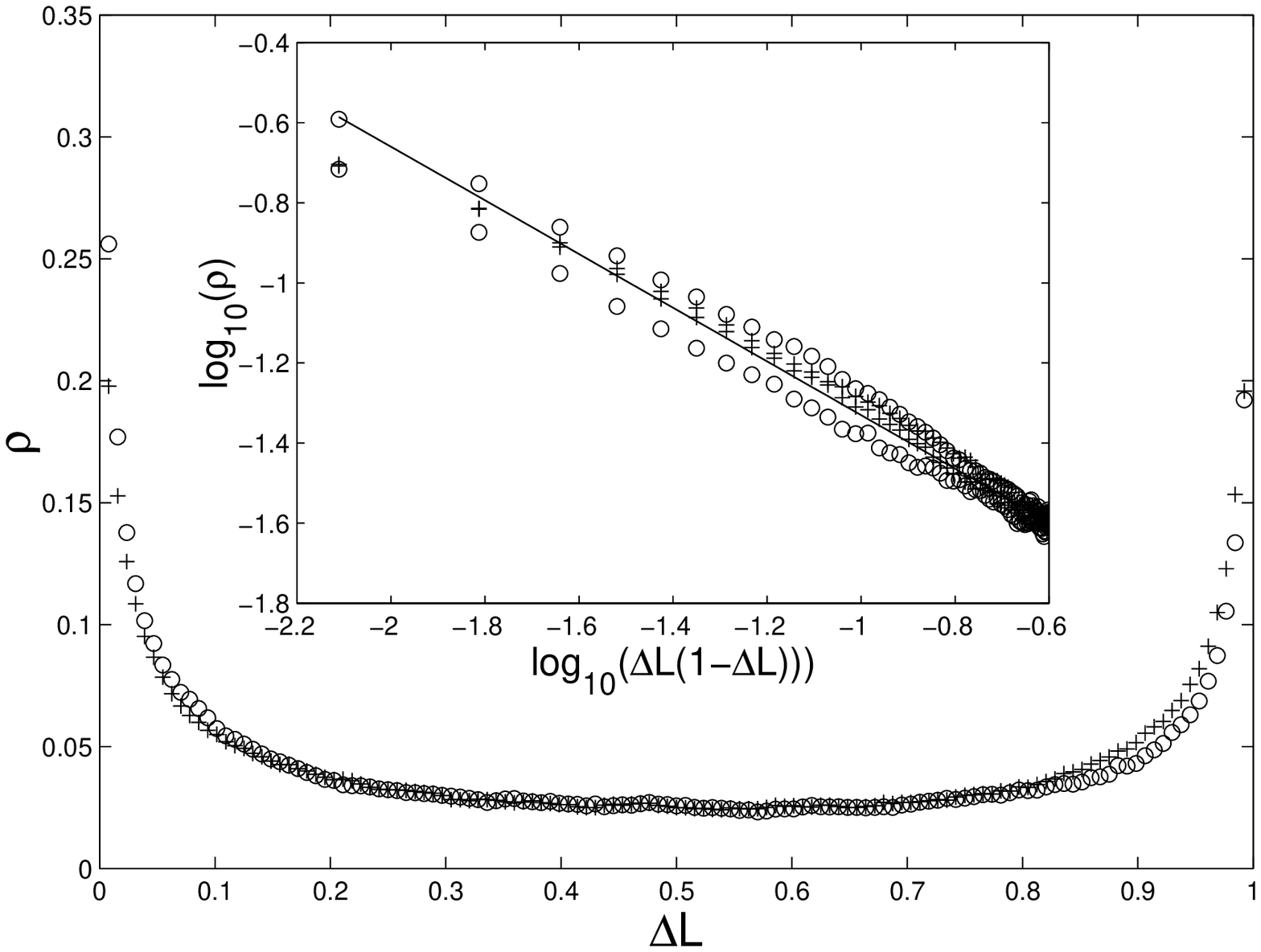}  
\caption{(a) AFM picture of the bead surface (b) density $\rho$ as a function 
of the relative difference $\Delta L$ with $L_2=128$. $o$ corresponds to row 
averaging and $+$ correspond to column averaging. Inset : $\log_{10}(\rho)$ 
as a function of $\log_{10}(\Delta L(1-\Delta L))$. The straight line 
corresponds to a fit using Eq.~\ref{e.4}}
\label{f.4}
\end{figure}

AFM frames are stored as $256\times 256$ matrices. $L_2$ is fixed to 128 and 
$\rho(\Delta L)$ has been determined for rows and columns averaging (see 
fig.~\ref{f.4}). In both case, we find $\zeta=0.35 \pm 0.05$, and therefore 
$1-\zeta \simeq 0.65$. This value is consistent with $\alpha_t=0.6$.

\section{Conclusion}

In summary, a series of electrical measurements has been performed to test 
the SAM predictions of large force chains rearrangements in a 2D metallic 
beads packing under small perturbations. As soon as the packing is heated or 
cooled, its conductance starts to decrease and to fluctuate. These 
fluctuations are intermittent and gather in bursts as predicted by the SAM. 
However, the waiting time distribution inside bursts are independent of beads 
network, perturbation intensity and external applied stress. Consequently, 
the origin of these fluctuations should be found in local microcontact 
rearrangements at each beads rather than collective vaults rearrangements, as 
suggested in \cite{b.12}. However, some information on these collective 
rearrangments
might lie in the deviation from a pure power-law distribution which is 
clearly observed on figure 4 when one increases the number of grains. These
collective effects might also be needed to account for the variation of the
exponent $\alpha_g$ with the geometry.
Work in this direction is underway.

\acknowledgments
We wish to thank F. Mangeant, E. Bonneville and M. Planelles for their 
participation to the data collection. We also acknowledge discussions with T. 
Baumberger, M. Bonetti, M. Cates, P. Evesque,  E. Kolb, J. Wittmer and  G. Zalczcer. 
We thank C. Gasquet and P. Meininger for technical support.
We also acknowledge F. Dubreuil for AFM frames of our beads. Ph. Claudin is currently supported by an Aly 
Kaufman postdoctoral fellowship.

\end{document}